\documentclass{article}

\usepackage{arxiv}

\usepackage{graphicx}
\usepackage{amsmath,amssymb,amsfonts}
\usepackage{algorithm}
\usepackage[noend]{algpseudocode}
\usepackage{textcomp}
\usepackage{tikz}
\usepackage{amsthm}
\usepackage{pifont}
\usepackage{hyperref}
\usepackage{multirow}
\usepackage{bm}
\usepackage{pgfplots}
\pgfplotsset{compat=1.18}
\usepackage{pgfplotstable}
\usepackage[table]{xcolor}
\usepackage{listings}

\newcommand*\circled[1]{%
  \tikz[baseline=(X.base)]{
    \node[draw, circle, fill=black, text=white, inner sep=0.5pt, minimum size=0.8em] (X) {\scriptsize\textbf{#1}};
  }%
}

\newcommand{\DM}{\text{$\Delta_m$}}
\newcommand{\secref}[1]{§\ref{#1}}

\title{TC-MIS: Maximal Independent Set on Tensor-Cores}

\author{
 Prajjwal Nijhara \\
  Department of Computer Science and Engineering\\
  Indian Institute of Technology Jodhpur\\
  Karwar, Rajasthan, India \\
  \texttt{d23cse005@iitj.ac.in} \\
   \And
Dip Sankar Banerjee \\
  Department of Computer Science and Engineering\\
  Indian Institute of Technology Jodhpur\\
  Karwar, Rajasthan, India \\
  \texttt{dipsankarb@iitj.ac.in} \\}
 
\begin{document}
\maketitle
\begin{abstract}
\textbf{Maximal Independent Set (MIS) in a graph is a fundamental problem with applications in resource allocation, scheduling, and network optimization. Although graphs are inherently unstructured and challenging for GPU parallelism due to irregular memory access and workload imbalance, specialized GPU algorithms have achieved good performance, processing million-vertex graphs in milliseconds. Modern GPUs are equipped with Tensor Cores (TCs), specialized units for matrix operations with 8-16$\times$ higher throughput than CUDA Cores (CCs), which are extensively used for ML, DL, and inference tasks but remain largely unexplored for graph algorithms. In this paper, we present TC-MIS, a TC-accelerated algorithm that reformulates key phases of MIS computation as sparse matrix-vector multiplication (SpMV). TC-MIS tiles the graph adjacency matrix and employs Warp Matrix Multiply-Accumulate (WMMA) operations to transform irregular graph traversal into regular, massively parallel computation. Our evaluation across TC-enabled microarchitectures (Ampere, Ada Lovelace, Hopper, Blackwell) demonstrates that TC-MIS achieves an average speedup of 2.84$\times$ on RTX A5000, 4.84$\times$ on L40S, 18.80$\times$ on H200 GPUs, and 5.20$\times$ on RTX 5080 with a maximum speedup of 44.38$\times$ on H200 GPU over state-of-the-art methods, while maintaining solution quality comparable to that obtained by established heuristics that produce near-maximum independent sets.}
\end{abstract}


\section{Introduction}

An independent set in a graph is a subset of vertices where no two vertices are adjacent, meaning there are no edges connecting any pair of vertices within the set. When an independent set is maximal, such that no additional vertex can be added to the set while maintaining the independence property, it is called a Maximal Independent Set (MIS). The largest possible independent set in a graph is known as the Maximum Independent Set. However, finding such a set requires comparing all possible independent sets, which makes the problem NP-hard~\cite{Garey}. Since any MIS satisfies the key properties of the maximum such set and can effectively solve underlying applications such as graph coloring~\cite{luby1986simple}, job scheduling~\cite{gertsbakh1978minimal}, structural analysis of protein~\cite{hobohm1992selection}, and network optimization~\cite{widgerson, tarjan1977finding}, MIS is often considered a practical alternative, with efficient heuristics available for computing it quickly~\cite{ghaffari2016improved, ghaffari2018improved}.

Sequential approaches to MIS computation, including brute-force enumeration, backtracking, and greedy heuristics, have been well studied in the literature~\cite{Garey, tsukiyama1977new, widgerson}. However, the rapid growth of large-scale graph datasets arising from real-world systems such as social, biological, and communication networks~\cite{girvan2002community, barabasi1999emergence, shen2022graph, newman2003structure} has exposed the scalability limitations of sequential algorithms. For graphs with millions or billions of edges, worst-case exponential complexity and per-iteration $\mathcal{O}(|V|+|E|)$ costs make even efficient heuristics impractical. This has motivated extensive research into parallel MIS algorithms, notably the randomized approaches of Karp and Widgerson~\cite{widgerson} and Luby~\cite{luby1985simple, luby1986simple}, which demonstrated effective parallelization through priority-based vertex selection and inspired subsequent multi-core CPU implementations.


Despite the success of multi-core CPU parallelization~\cite{tarjan, mcgregor2014graph}, graph algorithms pose significant challenges for GPU acceleration due to irregular memory access patterns, skewed vertex degrees, and data-dependent control flow. Nevertheless, several GPU-based MIS algorithms have demonstrated substantial performance improvements over CPU implementations~\cite{cusp, pannotia, irgl}. Among these, ECL-MIS by Burtscher et al.~\cite{Burtscher} represents the state of the art, achieving high performance through a degree-aware priority assignment scheme and a lock-free, deterministic vertex selection strategy operating directly on compressed sparse row (CSR) graph representations.



\vspace{-1.2em}
\subsection{Motivation}

GPUs are built for regular, data-parallel computation with predictable control flow and coalesced memory accesses. In contrast, graph algorithms are inherently irregular due to skewed degree distributions, variable-length adjacency lists, and data-dependent traversal patterns. These characteristics lead to load imbalance, warp divergence, irregular memory accesses, and frequent synchronization, all of which limit effective GPU utilization. Prior GPU graph algorithms mitigate these challenges through compressed sparse representations such as CSR, edge- or vertex-centric execution models, and lock-free, asynchronous updates. These techniques have enabled significant acceleration of graph workloads on GPUs, but they fundamentally operate within the constraints imposed by irregular neighbor traversal on CUDA Cores (CCs).

\begin{figure}[h]
    \centering
    \includegraphics[width=0.5\linewidth]{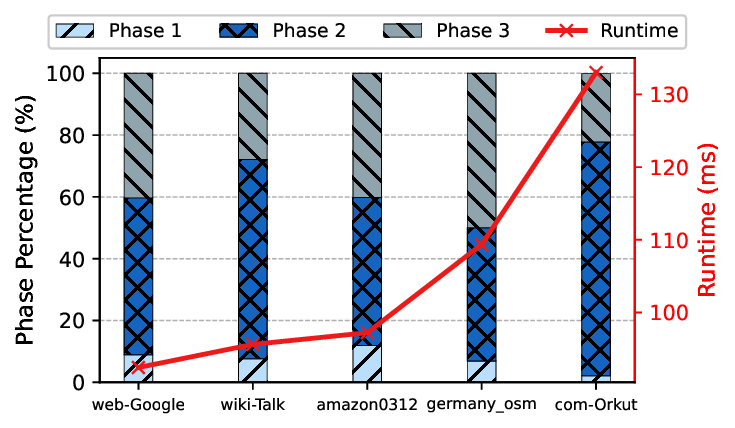}
    \caption{\textbf{Profiling of ECL-MIS~\cite{Burtscher}. The left Y-axis shows the percentage of time spent in each phase, while the right Y-axis shows the total MIS computation time on an L40s GPU.}}
    \label{fig:ecl-profiling}
\end{figure}

Similar problems arise when parallelizing MIS on GPUs, which are addressed by parallel implementations of Luby's algorithm. ECL-MIS~\cite{Burtscher}, built on this foundation, represents the state of the art, combining a Luby-style randomized algorithm with a degree-aware priority heuristic to achieve millisecond-scale performance on million-vertex graphs. However, our profiling reveals that ECL-MIS executes a three-phase algorithm in each iteration: \circled{1}~priority generation and neighbor maximum computation, \circled{2}~candidate selection and neighbor elimination, and \circled{3}~state update. As shown in Figure~\ref{fig:ecl-profiling}, Phase~\circled{2} dominates execution time, accounting for an average of 56.4\% of total runtime across diverse graph inputs. This phase relies on repeated traversal of variable-length neighbor lists and concurrent updates to vertex states, resulting in irregular memory access patterns and limited parallel efficiency. These observations motivate us to explore whether this dominant neighborhood processing step can be reformulated into a more regular computation that better aligns with modern GPU hardware capabilities.



Meanwhile, modern GPUs are equipped with specialized Tensor Cores (TCs), hardware units designed to accelerate matrix Warp Matrix Multiplication Accumulation (WMMA) operations with substantially higher throughput than conventional CCs. TCs are particularly effective for ML/DL workloads, which exhibit high arithmetic intensity, regular memory access patterns, and structured linear algebraic computations that map naturally to SpMM(V) operations. For MIS in particular, bottlenecks arise from the irregular traversal of adjacency lists and scattered memory accesses that are poorly suited for execution on TCs. Moreover, the CSR representation does not naturally decompose into the fixed $16 \times 16$ tiles required by WMMA operations. As a result, despite their widespread availability on modern GPUs, Tensor Cores remain largely unexplored for irregular graph algorithms such as MIS. These observations motivate the following questions: \textit{Can we re-engineer the existing parallel GPU-based MIS algorithms to leverage the massive computational throughput of TCs? More broadly, can Tensor Cores be utilized beyond traditional dense linear algebra to accelerate irregular, general-purpose graph computations?} 


\subsection{Contributions}
We propose TC-MIS, a TC-based GPU-first algorithm for computing MIS that operates asynchronously and produces deterministic results. Our contributions in this work are summarized as follows: (1) We reformulate the MIS computation as SpMV, enabling the use of WMMA operations on TCs. (2) We design a tiling and execution strategy that converts irregular graph traversal into structured computation compatible with execution on TCs, while preserving correctness and avoiding synchronization overheads associated with fine-grained atomic updates. (3) We conduct a comprehensive evaluation on multiple TC-enabled GPU microarchitectures (Ada Lovelace, Ampere, Hopper, Blackwell) and study the impact of different priority assignment heuristics on performance, demonstrating an average speedup of 18.80$\times$ over state-of-the-art GPU MIS implementations while maintaining solution quality comparable to established heuristics that produce near-maximum independent sets.

\section{Background}




Luby’s algorithm is the most widely adopted parallel algorithm for computing MIS. It is a randomized algorithm that achieves logarithmic expected time via randomized symmetry breaking. As illustrated in Algorithm~\ref{alg:luby-mis}, each iteration undergoes three phases. In the first phase (Line~3), every active vertex independently selects a random priority. In the second phase (Line~4), vertices compare their priorities with those of their neighbors, and vertices whose priorities are strictly larger than all adjacent active vertices are marked as candidates. In the third phase (Line~5), all candidate vertices are added to the MIS, and both the candidates and their neighbors are removed from the active set. This process repeats until no active vertices remain (Line~2). Luby’s algorithm serves as the foundation for many practical parallel MIS implementations on shared-memory and GPU architectures.









\begin{algorithm}
\caption{Luby’s Parallel MIS Algorithm}
\label{alg:luby-mis}
\begin{algorithmic}[1]
\Require Graph $G=(V,E)$
\Ensure MIS \text{\DM} $\subseteq V$

\State $\mathcal{A} \gets V$, \text{\DM} $\gets \emptyset$

\While{$\mathcal{A} \neq \emptyset$}

    \State \textbf{Phase~\circled{1}:} Assign random priorities $p(v)$ to all $v \in \mathcal{A}$

    \State \textbf{Phase~\circled{2}:} $C \gets \{\, v \in \mathcal{A} \mid p(v) > p(u),\ \forall u \in \mathcal{N}(v)\cap\mathcal{A} \,\}$

    \State \textbf{Phase~\circled{3}:} $\text{\DM} \gets \text{\DM} \cup C$, \;
    $\mathcal{A} \gets \mathcal{A} \setminus \big( C \cup \mathcal{N}(C) \big)$

\EndWhile

\State \Return \text{\DM}
\end{algorithmic}
\end{algorithm}

ECL-MIS builds~\cite{Burtscher} on the random-permutation variant of Luby’s algorithm, in which priorities are assigned first to induce a global ordering and are reused across iterations. Instead of using purely random priorities, ECL-MIS employs a degree-aware priority assignment that biases selection toward lower-degree vertices, which empirically reduces conflicts and accelerates convergence. Specifically, ECL-MIS computes priorities based on a monotonic function of vertex degree,

\begin{equation}
\label{eq:ecl-heuristic}
P(v) = \frac{\bar{d}}{\bar{d} + \deg(v) - \epsilon(v)}
\end{equation}

where $\deg(v)$ denotes the degree of $v$, $\bar{d}$ is the average graph degree, and $\epsilon(v)$ is a small randomized perturbation. In practice, this value is scaled and discretized to fit a compact integer representation suitable for GPU execution. This formulation preserves the probabilistic symmetry-breaking behavior of Luby’s algorithm while favoring vertices that are less likely to conflict. ECL-MIS further exploits massive GPU parallelism by implementing candidate selection and neighbor elimination in a lock-free, asynchronous manner using persistent kernels. 

\begin{figure*}
    \centering
    \includegraphics[width=\linewidth]{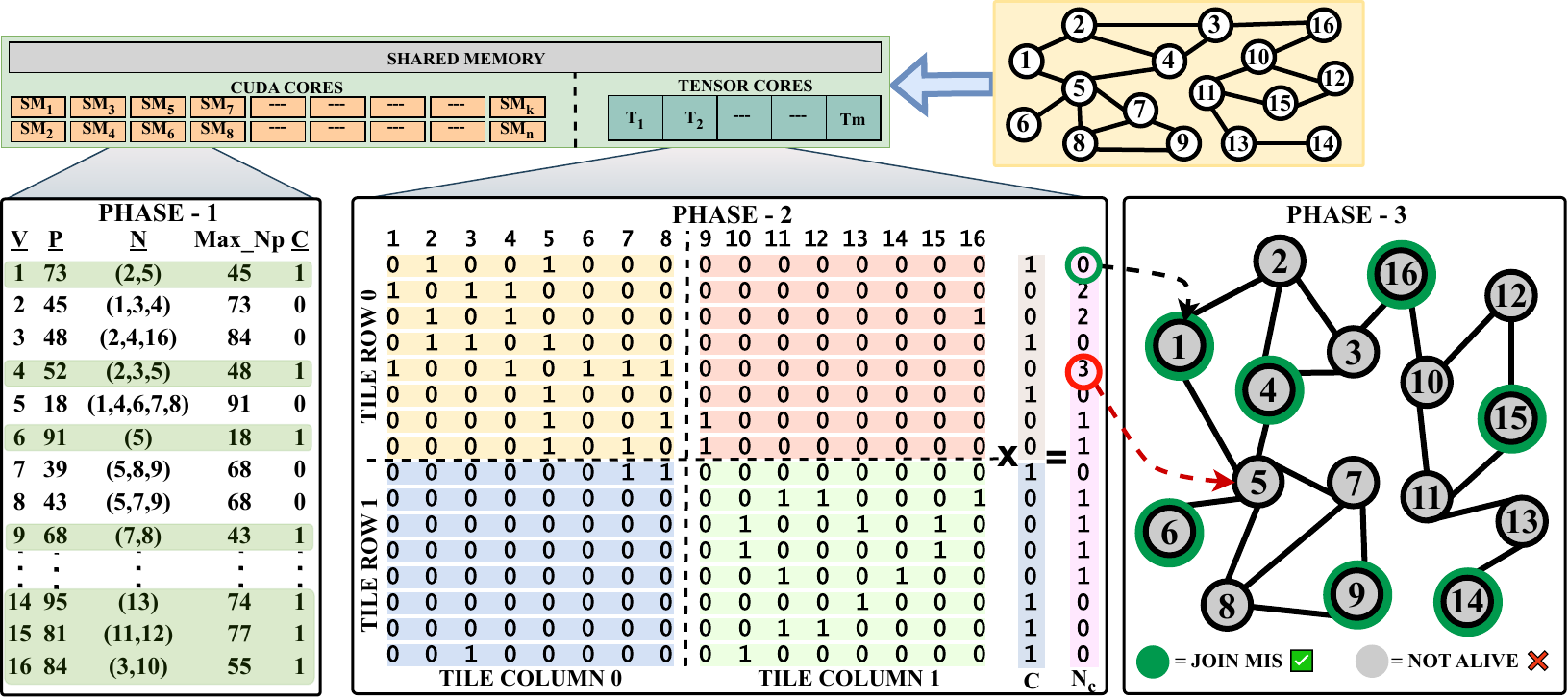}
    \caption{\textbf{Three-phase execution of TC-MIS illustrating heterogeneous GPU utilization. Phase~1 computes vertex priorities and identifies candidate vertices using CCs. Phase~2 performs block SpMV on an illustrative $8\times8$ tiled adjacency matrix using TCs to evaluate neighborhood interactions (shown for simplicity). Phase~3 updates vertex states based on the resulting vector using CCs, removing selected vertices and their neighbors from further consideration.}}
    \label{fig:tc-mis-example}
\end{figure*}

\subsection{Existing Works}
Early work on parallel MIS focused on randomized symmetry-breaking algorithms, beginning with Luby's seminal algorithms~\cite{luby1985simple, luby1986simple}. Subsequent work demonstrated that randomized approaches consistently outperform deterministic methods~\cite{alon1986fast, ghaffari2016improved, lenzen2011mis}. These ideas form the basis of GPU-based MIS implementations such as CUSP~\cite{cusp}, Pannotia~\cite{pannotia}, and IrGL~\cite{irgl}. Among them, ECL-MIS~\cite{Burtscher} represents the current state of the art, combining degree-aware priorities, asynchronous execution, and fine-grained conflict resolution to achieve millisecond-scale runtimes on large graphs. However, all existing GPU MIS implementations rely exclusively on CCs and remain limited by irregular memory access, load imbalance, and fine-grained edge-level processing.

Separately, linear-algebraic formulations have proven effective for accelerating graph analytics. CombBLAS~\cite{bulucc2011combinatorial, azad2021combinatorial} and the GraphBLAS standard~\cite{kepner2016mathematical, mattson2017graphblas} demonstrated how graph primitives can be expressed using sparse matrix operations. Extensive work on optimizing SpMV and SpGEMM on GPUs~\cite{aliaga2022compression, anzt2020load, chen2022fgspmspv, li2023haspmv, niu2021tilespmv} shows that block-based formulations can significantly improve hardware utilization. TCs, originally designed for dense matrix operations in machine learning~\cite{choquette2021nvidia}, have recently been applied to stencil computations, FFTs, and sparse linear solvers~\cite{chen2024convstencil, li2021tcfft, haidar2018harnessing}. Their use in graph algorithms, however, remains limited, with only a small number of exploratory studies on BFS, triangle counting, and shortest paths~\cite{niu2025berrybees, chen2025tot, firoz2020feasibility}. To the best of our knowledge, no prior work reformulates MIS to exploit TCs. This gap motivates our work, which maps the core neighbor-counting step of MIS to a TC-accelerated linear-algebraic kernel.

\section{TC-MIS Implementation}

TC-MIS reformulates Luby’s iterative framework (Algorithm~\ref{alg:luby-mis}) to leverage the heterogeneity of modern GPUs. While existing GPU implementations, such as ECL-MIS~\cite{Burtscher}, CUSP~\cite{cusp}, Pannotia~\cite{pannotia}, and IrGL~\cite{irgl}, rely solely on CCs, TC-MIS is guided by the observation that the dominant cost of parallel MIS lies in repeated neighborhood interactions that determine candidate conflicts. Rather than executing these interactions through irregular adjacency-list traversals, TC-MIS re-engineers them to expose structured parallelism, enabling the most expensive step of MIS to be expressed as a SpMV operation and transparently executed on Tensor Cores. This shift allows TC-MIS to exploit TC throughput while preserving the algorithmic semantics of Luby-style MIS.

\subsection{Base Algorithm}
\label{ssec:tc-mis-base}
We present TC-MIS in Algorithm~\ref{alg:tc-mis}, which exploits heterogeneous GPU resources by formulating the computationally expensive steps as TC-accelerated SpMV.

\textbf{Phase \circled{1}: Priority Assignment and Candidate Generation.}
Vertex priorities are initialized using a heuristic function similar to Equation~\eqref{eq:ecl-heuristic}, and all vertices are initially marked as alive $\mathcal{A}$ (Lines~1--4). For each vertex $v$, we compute in parallel the maximum priority among its neighbors, producing a vector $\text{Max\_Np}$ such that  $\text{Max\_Np}(v) \gets \max_{u \in \mathcal{N}(v) \cap \mathcal{A}} P(u)$ (Line~5). Using this information, we generate the candidate vector $C$ in parallel, where $C(v) = 1$ if $P(v) > \text{Max\_Np}(v)$ and $C(v) = 0$ otherwise (Line~6). As illustrated in Figure~\ref{fig:tc-mis-example}, for vertex $v=1$, we have $P(1)=73$ and $\text{Max\_Np}(1)=45$, and therefore $C(1)=1$, indicating that vertex~1 is selected as a candidate.

\textbf{Phase \circled{2}: Neighbor Counting via TCs.}
Phase~\circled{2} formulates neighbor counting as SpMV executed using tile-level operations on TCs. We represent the graph as a block-tiled adjacency matrix $A$ composed of $16 \times 16$ tiles; for clarity, Figure~\ref{fig:tc-mis-example} illustrates this process using $8 \times 8$ tiles. Multiplying $A$ with the candidate vector $C$ produces the neighbor count vector $N_c$, where $N_c(v)$ denotes the number of candidate neighbors of vertex $v$ (Lines~7–8). This formulation replaces irregular adjacency-list traversals with regular tile-based computation, enabling efficient parallel execution via WMMA and significantly accelerating Phase~\circled{2}.

\textbf{Phase~\circled{3}: Lock-Free Vertex State Update.}
Using the candidate vector $C$ and the neighbor count vector $N_c$, vertex states are updated in parallel according to the following rules:
\begin{enumerate}
\item If $C(v) = 1$, vertex $v$ is selected into \text{\DM} and marked inactive.
\item If $C(v) = 0$ and $N_c(v) > 0$, vertex $v$ is adjacent to at least one candidate and is therefore marked inactive.
\item If $C(v) = 0$ and $N_c(v) = 0$, vertex $v$ remains active and is considered in the next iteration (Lines~9–11).
\end{enumerate}

\begin{algorithm}
\caption{TC-MIS}
\label{alg:tc-mis}
\begin{algorithmic}[1]
\Require Graph $G = (V,E)$
\Ensure MIS \text{\DM} $\subseteq V$

\State $\mathcal{A} \gets V$, $\mathcal{M} \gets \emptyset$
\While{$\mathcal{A} \neq \emptyset$}

    \Statex \textbf{Phase~\circled{1}: Priority Assignment and Candidate Generation}
    \ForAll{$v \in \mathcal{A}$ \textbf{in parallel}}
        \State $P(v) \gets$ priority heuristic
        \State $\text{Max\_Np}(v) \gets \max_{u \in \mathcal{N}(v) \cap \mathcal{A}} P(u)$
        \State $C(v) \gets \mathbb{1}\big[P(v) > \text{Max\_Np}(v)\big]$
    \EndFor

    \Statex \textbf{Phase~\circled{2}: Neighbor Counting via Tensor Cores}
    \State Represent $G$ as tiled adjacency matrix $A$
    \State $N_c \gets A \times C$ \Comment{SpMV using WMMA}

    \Statex \textbf{Phase~\circled{3}: Lock-Free Vertex State Update}
    \ForAll{$v \in \mathcal{A}$ \textbf{in parallel}}
        \State \text{\DM} $\gets \text{\DM} \cup \{v \mid C(v)=1\}$
        \State $\mathcal{A} \gets \mathcal{A} \setminus \{v \mid C(v)=1 \lor N_c(v)>0\}$
    \EndFor

\EndWhile
\State \Return \text{\DM}
\end{algorithmic}
\end{algorithm}

Neighbor elimination is handled implicitly through $N_c$, avoiding explicit updates to adjacent vertices. This phase is lock-free by construction, as each vertex independently reads $C(v)$ and $N_c(v)$ and updates only its own state, without requiring atomic operations or synchronization. As illustrated in Figure~\ref{fig:tc-mis-example}, vertex $v = 1$ satisfies $C(1)=1$ and $N_c(1)=0$ and is therefore added to \text{\DM}, while vertex $5$ satisfies $C(5)=0$ and $N_c(5)>0$ and is consequently marked inactive.

\begin{lstlisting}[
caption={\textbf{TC-MIS: TC execution via WMMA API}},
label={lst:tc-mis-base},
language=C++,
basicstyle=\scriptsize\ttfamily]
// Phase 2: TC accelerated SpMV via WMMA
__global__ void phase2_tensorcore_spmv(
    const half* A_tiles,        // non-zero 16x16 adjacency tiles
    const half* C_tiles,        // tiled candidate vector
    int* Nc)                    // neighbor-interaction predicate
{
    using namespace nvcuda::wmma;

    fragment<matrix_a,16,16,16,half,row_major>  A_frag;
    fragment<matrix_b,16,16,16,half,col_major> C_frag;
    fragment<accumulator,16,16,16,float>       Acc_frag;

    fill_fragment(Acc_frag, 0.0f);

    // Load one adjacency tile and corresponding candidate segment
    load_matrix_sync(A_frag, A_tiles, 16);
    load_matrix_sync(C_frag, C_tiles, 16);

    // WMMA execution on Tensor Cores
    mma_sync(Acc_frag, A_frag, C_frag, Acc_frag);

    // Reduce tile result into per-vertex interaction predicate
    accumulate_rows(Acc_frag, Nc);
}

\end{lstlisting}

\subsection{Implementation Details}
\label{ssec:tc-mis-optimizations}
Listing~\ref{lst:tc-mis-base} presents the TC implementation of Phase~\circled{2} in TC-MIS, focusing on design choices that enable efficient TC execution for irregular graph workloads.

Phase~\circled{1} is implemented as a CUDA kernel that produces the boolean candidate vector $C$, where $C(v)=1$ indicates that vertex $v$ locally dominates its active neighborhood. This kernel executes entirely on CCs and returns $C$ as its sole output. The computation involves irregular memory access and control divergence, making it ill-suited for TC execution. Accordingly, Phase~\circled{1} is treated as a preprocessing step that prepares a compact, structured input for subsequent TC-accelerated computation.

Phase~\circled{2} constitutes the computational core of TC-MIS. The goal is to determine, for each active vertex, whether it is adjacent to at least one candidate vertex. Rather than performing edge-centric traversals with fine-grained atomic updates, we formulate this operation as an SpMV, $N_c = A \times C$, where $A$ is the adjacency matrix and $C$ is the candidate vector produced in Phase~\circled{1}. To enable execution on TCs, $A$ is represented as a collection of non-zero $16\times16$ tiles stored in a compressed block format (Line~9), and $C$ is packed into a compatible tiled layout (Line~10). Each non-zero tile is mapped to an independent CUDA thread block and processed by a single warp using WMMA primitives (Lines~16-17). At execution time, warps are scheduled onto SMs, where TCs execute the corresponding \texttt{mma\_sync} instructions (Line~20). Because the total number of non-zero tiles typically far exceeds the number of resident warps and available TCs, tiles are executed in multiple waves: only a subset of tiles is active concurrently, while the remainder are queued and scheduled as hardware resources become available. This temporal reuse of resources allows TC-MIS to process arbitrarily large graphs despite the fixed tile size supported by WMMA.



Within each active warp, threads cooperatively load the tile operands, invoke the WMMA operation, and accumulate partial results in FP32. Accumulation precision is sufficient because each tile contributes at most 16 to any output entry, and the final result is used only to evaluate the predicate $N_c(v) > 0$. To avoid unnecessary TC execution, tiles whose corresponding segment of $C$ contains no non-zero entries are skipped entirely. This early-exit mechanism becomes increasingly effective in later iterations as the candidate set shrinks, reducing the number of WMMA tasks scheduled for execution.


\paragraph{Memory Footprint and Representation Trade-offs.}
Representing the graph as a tiled adjacency matrix incurs higher memory overhead. We reduce this cost by storing only non-zero tiles and using compact data types for tile entries and candidate vectors. This design intentionally trades space efficiency for computational regularity that enables TC acceleration of the dominant kernel. In practice, we found the memory footprint remains manageable for large sparse graphs due to the low density of non-zero tiles (See~\secref{sec:eval}).

\paragraph{Synchronization and Lock-Free Updates.}
TC-MIS minimizes synchronization overhead across all phases. In Phase~\circled{2}, atomic operations are used only at tile granularity (once per row per tile) rather than per-edge, significantly reducing contention. Phase~\circled{3} is lock-free by construction: each thread independently updates only its own vertex state based on local reads of the candidate and neighbor count vectors. Atomic operations are limited to maintaining global counters and are not on the critical path of computation.

\subsection{Heuristic Functions for Priority Assignment}
\label{ssec:heuristics}
As Phase~\circled{1} of Algorithm~\ref{alg:tc-mis} assigns priorities to vertices using a priority heuristic, the choice of heuristic directly controls which vertices are favored as candidates for inclusion in the MIS. Consequently, the priority assignment significantly influences both the solution quality and the overall runtime. To balance these objectives for TC-MIS, we implement three priority heuristics to improve MIS quality while keeping computational overhead low and execution time manageable.

\textbf{H1 (Random Priority)} assigns priorities as $P(v) = hash(v)$. H1 minimizes computation and maximizes parallelism, but is expected to yield low MIS cardinality due to the absence of structural bias.

\textbf{H2 (Degree-Aware Priority)} incorporates local structure by scaling priorities similar to Equation~\ref{eq:ecl-heuristic}. H2 favors low-degree vertices with the aim of improving MIS quality over H1. H2 is effective in ECL because strict, sequential neighbor comparisons exploit degree-aware priorities without requiring explicit conflict resolution. However, when implemented in TC-MIS, H2 prematurely eliminates many viable vertices under parallel, tile-based execution, motivating the use of explicit conflict resolution to preserve solution quality.

\textbf{H3 (Degree-Aware Priority with Conflict Resolution)} eliminates greedy removal of vertices from $\mathcal{A}$ and instead enforces an ordered resolution. Vertices first remain in $\mathcal{A}$ while conflicts are resolved on a pending set to produce a conflict-free candidate vector $C$. Only after $C$ is finalized are vertex states updated, and removal is applied according to $\mathcal{A} \leftarrow \mathcal{A} \setminus \{v \mid C(v)=1 \lor N_c(v)>0\}$. This logical ordering, \emph{Alive set} $\rightarrow$ \emph{conflict resolution} $\rightarrow$ \emph{candidate finalization} $\rightarrow$ \emph{state update}, avoids premature eliminations while preserving parallelism, with lower overhead.

\section{Evaluation}
\label{sec:eval}
In this section, we evaluate TC-MIS by comparing it against ECL-MIS, the state-of-the-art GPU implementation of Luby-style MIS. Prior work has shown that ECL-MIS consistently outperforms earlier GPU-based implementations, such as CUSP, Pannotia, and IrGL, in both runtime and solution quality. We therefore use ECL-MIS as a strong baseline to assess the effectiveness of TC-MIS.

\subsection{Experimental Setup}
\subsubsection{System}
\label{ssec:system}
We evaluate TC-MIS on four TC-equipped NVIDIA GPUs spanning multiple architectural generations to assess robustness. The NVIDIA RTX A5000 (Ampere microarchitecture) is a workstation-class GPU with 64 SMs and 8{,}192 CCs, equipped with 24 GB of GDDR6 memory and a 384-bit memory interface. The NVIDIA L40S (Ada Lovelace microarchitecture) with 142 SMs and 18{,}176 CCs, equipped with 48 GB of GDDR6 memory and a 384-bit memory interface, offering high compute throughput with moderate memory bandwidth. The NVIDIA RTX 5080 (Blackwell microarchitecture) represents a high-end consumer-class GPU, equipped with 10{,}752 CCs, 16 GB of GDDR7 memory on a 256-bit interface. The NVIDIA H200 (Hopper microarchitecture) represents a data center-class accelerator with 132 SMs and 16{,}896 CCs, equipped with 141 GB of HBM3 memory and a 6144-bit memory interface, providing substantially higher memory bandwidth and compute capability. All experiments were conducted using CUDA 12.4 and the latest NVIDIA drivers.


\begin{table}[h]
\centering
\caption{\textbf{Graphs used in our experiments from the \textit{SuiteSparse Matrix Collection}~\cite{tamu-mtx}.}}
\label{tab:graph_properties}
\begin{tabular}{|c||c||c||c||c|}
    \hline
    \textbf{S. No.} & \textbf{Graph} & \textbf{$|V|$ (M)} & \textbf{$|E|$ (M)} & \textbf{$|E|/|V|$} \\ \hline
    G1 & amazon0302        & 0.26 & 2.35   & 9.0  \\ \hline
    G2 & roadNet-PA        & 1.09 & 2.93   & 2.7  \\ \hline
    G3 & delaunay\_n19     & 0.52 & 3.00   & 5.7  \\ \hline
    G4 & wiki-Talk         & 2.39 & 9.54   & 4.0  \\ \hline
    G5 & web-Google        & 0.92 & 9.70   & 10.6 \\ \hline
    G6 & web-BerkStan      & 0.69 & 14.44  & 21.0 \\ \hline
    G7 & soc-LiveJournal1  & 4.85 & 68.99  & 14.2 \\ \hline
    G8 & kron\_g500-logn21 & 2.10 & 182.08 & 86.8 \\ \hline
\end{tabular}
\end{table}

\subsubsection{Datasets}
\label{ssec:datasets}
We use both sparse and dense graphs, ranging from 0.26M to 4.85M vertices and 2.45M to 182.08M edges, as summarized in Table~\ref{tab:graph_properties}. All graphs are undirected and unweighted, as the MIS problem does not require edge directionality or weights. This mix of sparse and dense graphs enables the evaluation of TC-MIS across diverse structural characteristics and workload intensities. Sparse graphs emphasize irregular memory access and control divergence, whereas dense graphs increase arithmetic intensity and expose opportunities for TC utilization through block-level regularity.


\subsection{Performance of TC-MIS}
\subsubsection{Solution Quality of TC-MIS Heuristics}

Before comparing the runtime performance of TC-MIS with ECL-MIS, we first evaluate solution quality using MIS cardinality obtained from different priority heuristics. As discussed in \secref{ssec:heuristics}, the degree-aware priority heuristic H2, originally proposed for ECL-MIS, is not directly suitable for TC-MIS due to differences in execution semantics. TC-MIS relies on tensor-centric and warp-synchronous execution, which requires a globally consistent priority flow, whereas directly applying degree-based heuristics can introduce conflicts during parallel updates. We therefore compare H1, H2, and H3 on TC-MIS against H2 on ECL-MIS. Since MIS cardinality is an algorithmic property independent of GPU microarchitecture, all experiments are conducted on an NVIDIA H200 GPU.

\begin{figure}[h]
    \centering
    \includegraphics[width=0.5\linewidth]{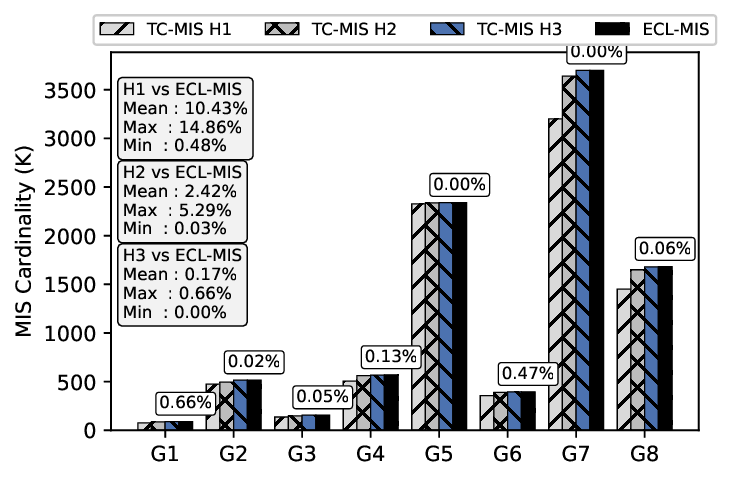}
    \caption{\textbf{MIS cardinality comparing the solution quality of TC-MIS priority heuristics (H1-H3) against the degree-aware heuristic of ECL-MIS. All experiments were conducted on an NVIDIA H200 GPU.}}
    \label{fig:solution-quality}
\end{figure}

\begin{figure*}
    \centering
    \includegraphics[width=\linewidth]{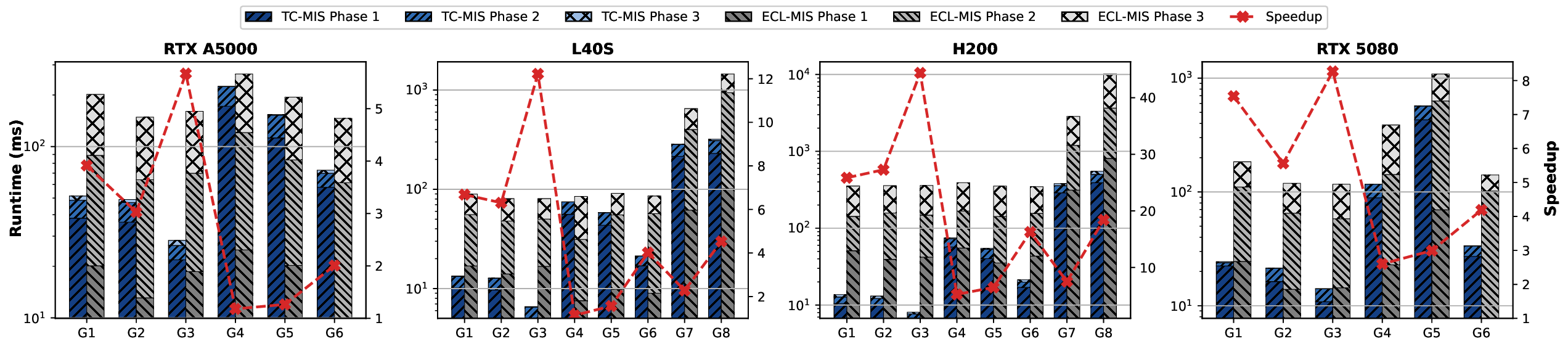}
    \caption{\textbf{Runtime comparison of TC-MIS and ECL-MIS across different GPUs (RTX A5000, L40S, H200, and RTX 5080). The left y-axis shows runtime (ms), and the right y-axis shows the speedup of TC-MIS over ECL-MIS. Graphs G7 and G8 are omitted for RTX A5000 and RTX 5080 because only graphs up to G6 fit in GPU memory for both implementations.}}
    \label{fig:runtime}
\end{figure*}

As shown in Figure~\ref{fig:solution-quality}, H1 yields the weakest solutions, with an average deviation of 10.43\% relative to ECL-MIS. This is expected given H1's purely random priority assignment without degree awareness. H2, which incorporates degree information, reduces the average deviation to 2.42\% but remains suboptimal. The deviation persists because H2 applies global degree-based ordering without respecting tile-level locality in the tensorized matrix representation, which results in priority inversions within tiles during warp-synchronous execution, particularly on graphs with skewed degree distributions. In contrast, H3 closely matches ECL-MIS's solution quality, achieving an average deviation of only 0.17\%. H3 achieves no deviation on graphs G5 and G7, where local degree ordering within tiles naturally aligns with optimal vertex selection patterns. These results demonstrate that H3 on TC-MIS preserves high solution quality, providing a fair basis for runtime comparison with ECL-MIS.

\subsubsection{Runtime Performance of TC-MIS}
For the runtime comparison between TC-MIS and ECL-MIS, we evaluate both implementations on the graphs listed in Table~\ref{tab:graph_properties} across four GPU architectures. Figure~\ref{fig:runtime} presents the runtime and speedup plots for RTX A5000, L40S, H200, and RTX 5080. Across all graphs, TC-MIS consistently outperforms ECL-MIS, with the magnitude of improvement increasing with newer GPU generations. On the RTX A5000, the average speedup over graphs G1--G6 is $2.84\times$; on the RTX 5080, it reaches $5.16\times$; on the L40S, the average speedup across G1--G8 increases to $4.84\times$; and on the H200, the average speedup further rises to $18.80\times$. This monotonic increase reflects the architectural scaling of TCs, which directly accelerates the WMMA-based SpMV kernel in Phase 2 of TC-MIS, while ECL-MIS remains constrained by CC execution throughout all three phases. Moving from L40S to H200 increases the average speedup from $4.84\times$ to $18.80\times$, a $3.88\times$ multiplicative gain, highlighting how Hopper's TC throughput and memory system amplify TC-MIS much more strongly than CC-based graph processing.

Speedup varies systematically with graph structure and density. The largest gains occur on sparse graphs such as G3 ($|E|/|V|=5.7$), which achieves the highest speedup across all GPUs (RTX A5000: $5.67\times$, RTX 5080: $8.27\times$, L40S: $12.22\times$, H200: $44.38\times$). G3's low density and regular structure produce highly sparse tiled adjacency matrices, allowing TC-MIS to process the SpMV in Phase 2 with near-peak TC utilization while minimizing memory traffic. On H200, TC-MIS on G3 spends 83.1\% of runtime in Phase 1, only 11.9\% in Phase 2 (TC-accelerated SpMV), and 5.0\% in Phase 3, whereas ECL-MIS distributes time as 11.7\%, 29.5\%, and 58.8\% across its phases. In contrast, G4 ($|E|/|V|=4.0$) exhibits the lowest speedup across all GPUs (RTX A5000: $1.18\times$, RTX 5080: $2.80\times$, L40S: $1.14\times$, H200: $5.23\times$) due to its highly skewed power-law degree distribution. Hub vertices create dense matrix rows that prevent effective tile skipping and generate irregular memory access patterns. Similarly, denser graphs like G6 ($|E|/|V|=21.0$) and G8 ($|E|/|V|=86.8$) achieve moderate speedups of $16.26\times$ and $18.40\times$ on H200, as high edge density reduces tile sparsity.

A critical factor driving TC-MIS's performance is the significant reduction in Phase 2 execution time through TC acceleration of the SpMV operation. On the RTX 5080 for G1, TC-MIS completes Phase 2 in just 6.3\% of total runtime versus ECL-MIS's 46.8\%, representing a $7.43\times$ reduction. On H200 for G3, Phase 2 accounts for only 11.9\% in TC-MIS compared to 29.5\% in ECL-MIS, a $2.48\times$ improvement. Across all GPUs and graphs, TC-MIS maintains Phase 2 times between 6.3\% and 26.2\%, while ECL-MIS ranges from 25.9\% to 66.7\%. This consistent advantage stems from replacing fine-grained edge-level operations with block-level WMMA execution, and becomes increasingly pronounced on newer architectures with higher TC-to-CC performance ratios.
\section{Conclusion}
In this paper, we present TC-MIS, a TC-accelerated implementation of Luby’s MIS algorithm that reformulates computationally intensive steps as tiled WMMA-based SpMV operations to map irregular graph computations onto TCs. We combine degree-aware priority heuristics with conflict-aware resolution to maintain MIS quality close to ECL-MIS while achieving consistently lower runtimes. Our evaluation on Ampere, Ada, Hopper, and Blackwell GPUs shows that TC-MIS scales with TC throughput, achieving an average speedup of 2.84$\times$ on RTX A5000, 4.84$\times$ on L40s, and 18.80$\times$ on H200 GPUs. This confirms that co-designing graph algorithms with TC execution delivers substantial performance gains beyond what conventional GPU kernels can achieve. In the future, we plan to optimize TC-MIS memory by implementing data structures common to both CCs and TCs, enabling the loading of billion-scale graphs into GPU memory. We also plan to extend our work to a dynamic setting where the graph structure changes frequently due to edge/vertex insertions/deletions.

\bibliographystyle{unsrt}  
\bibliography{refs}  

\end{document}